\documentclass[twocolumn]{aastex63}
\usepackage{CJK}
\usepackage{amsmath}
\usepackage{multirow}
\usepackage{graphicx}  
\usepackage{float}  
\usepackage{subfigure}  

\received{...}
\revised{...}
\accepted{...}

\shorttitle{Mass ratio distribution}
\shortauthors{Li et al.}

\begin{document}
\begin{CJK*}{UTF8}{gbsn}

\title{Divergence in mass ratio distributions between low-mass and high-mass coalescing binary black holes}

\author[0000-0001-5087-9613]{Yin-Jie Li \textsuperscript{*}（李银杰）}
\email{ * Contributed equally.}
\affiliation{Key Laboratory of Dark Matter and Space Astronomy, Purple Mountain Observatory, Chinese Academy of Sciences, Nanjing 210023, People's Republic of China}
\affiliation{School of Astronomy and Space Science, University of Science and Technology of China, Hefei, Anhui 230026, People's Republic of China}

\author[0000-0001-9626-9319]{Yuan-Zhu Wang \textsuperscript{*}（王远瞩）}
\affiliation{Key Laboratory of Dark Matter and Space Astronomy, Purple Mountain Observatory, Chinese Academy of Sciences, Nanjing 210023, People's Republic of China}

\author[0000-0001-9120-7733]{Shao-Peng Tang（唐少鹏）}
\affiliation{Key Laboratory of Dark Matter and Space Astronomy, Purple Mountain Observatory, Chinese Academy of Sciences, Nanjing 210023, People's Republic of China}
\affiliation{School of Astronomy and Space Science, University of Science and Technology of China, Hefei, Anhui 230026, People's Republic of China}

\author[0000-0003-4891-3186]{Qiang Yuan（袁强）}
\affiliation{Key Laboratory of Dark Matter and Space Astronomy, Purple Mountain Observatory, Chinese Academy of Sciences, Nanjing 210023, People's Republic of China}
\affiliation{School of Astronomy and Space Science, University of Science and Technology of China, Hefei, Anhui 230026, People's Republic of China}

\author[0000-0002-8966-6911]{Yi-Zhong Fan（范一中）}
\affiliation{Key Laboratory of Dark Matter and Space Astronomy, Purple Mountain Observatory, Chinese Academy of Sciences, Nanjing 210023, People's Republic of China}
\affiliation{School of Astronomy and Space Science, University of Science and Technology of China, Hefei, Anhui 230026, People's Republic of China}
\email{The corresponding author: yzfan@pmo.ac.cn (Y.Z.F)}

\author[0000-0002-9758-5476]{Da-Ming Wei（韦大明）}
\affiliation{Key Laboratory of Dark Matter and Space Astronomy, Purple Mountain Observatory, Chinese Academy of Sciences, Nanjing 210023, People's Republic of China}
\affiliation{School of Astronomy and Space Science, University of Science and Technology of China, Hefei, Anhui 230026, People's Republic of China}

\begin{abstract}
Coalescing binary black hole (BBH) systems are likely formed via several channels, and it is challenging to understand their formation / evolutionary processes. Some features in the mass function of the primary components ($m_1$), such as the distinct Gaussian-like peak located at $\sim 34M_\odot$, have been previously found. 
In this work, we investigate the possible dependence of the mass ratio ($q=m_2/m_1$) distribution on the primary mass. We find a Bayesian odds ratio of 18.1 in favor of divergence in the mass ratio distributions between the low- and high-mass ranges over an invariable mass ratio distribution. The BBHs with $m_1\gtrsim29M_{\odot}$ have a stronger preference to be symmetric compared to those with $m_1\lesssim29M_{\odot}$ at a 97.6\% credible level.
Additionally, we find mild evidence that the BBHs with $m_1$ located in the Gaussian-like peak have a mass ratio distribution different from that of other BBHs.
Our findings may be in favor of some formation channels, such as the chemically homogeneous evolution and the dynamical assembly in globular clusters/nuclear star clusters, which are more likely to provide symmetric BBHs in the high-mass range.
\end{abstract}

\keywords{}

\section{Introduction} \label{sec:intro}
The first successful detection of a gravitational-wave (GW) signal from a coalescing binary black hole (BBH) on 2015 September 14 \citep{2016PhRvL.116f1102A} has brought about the era of GW astronomy. Very recently, the LIGO-Virgo-KAGRA Collaborations (LVKC) reported the second part of the GW events detected in the third observing run (O3) \citep{2021arXiv211103606T}, and to date, nearly 100 detections have been reported \citep{2019PhRvX...9c1040A,2021PhRvX..11b1053A,2021arXiv210801045T,2021arXiv211103606T}. The number of detections may even reach 1000 once the GW detector network is observing in its design sensitivity \citep{2018LRR....21....3A}. 
However, the origins of these compact objects remain uncertain. Several evolutionary channels have been proposed \citep[see Refs.][for recent reviews]{Mapelli2020,2021NatAs...5..749G}, including, for instance, the isolated binary evolution and dynamical capture. These formation channels will leave an imprint on the properties of the compact binary population \citep{2018PhRvD..98h3017T,2019MNRAS.482.2991A}.
Therefore, studying the rapidly increasing population of GW events enables us to investigate how compact binaries form. Various studies that employed some analytical models \citep[e.g.,][]{2017PhRvD..96b3012T,2019ApJ...882L..24A,2021ApJ...913L...7A,2021arXiv211103634T} or some nonparametric approaches \citep[e.g.,][]{2021ApJ...917...33L,2021ApJ...913L..19T,2022ApJ...928..155T} were carried out, and some formation / evolutionary processes of the compact binaries are being revealed \citep[e.g.,][]{2021ApJ...913...42W,2021ApJ...915L..35K,2021ApJ...921L..15G,2021ApJ...907L..24S,2021ApJ...916L..16B,2021ApJ...922....3T,2022MNRAS.511.5797M,2021ApJ...923...97L,2021arXiv211010838W}.
 
The mass ratio may also carry some information about the formation and evolution mechanism of the BBHs.
For instance, \citet{2020ApJ...891L..27F} found that the two objects in the merging BBHs are more likely to be of comparable mass rather than randomly paired, consistent with the predictions of some formation channels 
\citep{2015ApJ...806..263D,2016MNRAS.458.3075A,2016PhRvD..93h4029R,2016MNRAS.458.2634M,2016A&A...588A..50M,2022PhR...955....1M}.
In particular, \citet{2016MNRAS.458.2634M} (see also \citep{2016MNRAS.460.3545D,2016A&A...588A..50M}) proposed a route toward merging massive BHs, i.e., the chemically homogeneous evolution, such that BBHs that form through it are expected to be most likely equal-mass components because the binary systems were in contact (shared mass) on the main sequence, before disengaging during subsequent phases of the chemically homogeneous evolution. 
The traditional isolated evolution channel, i.e., the common-envelope evolution, is also predicted to produce BBHs with comparable mass components. 
Anyhow, the chemically homogeneous evolution can only produce massive binaries with a total BH mass above $\sim 55 M_{\odot}$ \citep{2016MNRAS.458.2634M,2016A&A...588A..50M}, which exceeds the majority of the mass range for the common-envelope evolution, and the common-envelope evolution preference for equal-mass systems is less extreme than that by the chemically homogeneous evolution.
Additionally, the dynamical assembly, such as in globular clusters and nuclear star clusters, may also have a strong preference for symmetric masses \citep[see, e.g.,][and their references]{2016PhRvD..93h4029R,2017MNRAS.467..524B,2019MNRAS.486.5008A,2021ApJ...910..152Z}, and in such channels heavier BHs are more likely to merge \citep[see ][for a recent review]{2022PhR...955....1M}.
Therefore, the mass ratio distribution of BHs may be dependent on the primary mass.

In this Letter, we perform a hierarchical Bayesian inference to explore the features of the mass ratio distribution in the BBH populations. The work is organized as follows: In Section \ref{sec:method}, we introduce the data and the models used for inference, and in Section \ref{sec:result}, we present the results. We make our conclusion and discussion in Section \ref{sec:discussion}.
 
\section{Method} \label{sec:method}

\subsection{Selected events}\label{method_data}
Our analysis focuses on the GW data of BBHs reported in the Gravitational-wave Transient Catalog 3 (GWTC-3; \citep{2021arXiv211103606T}). 
To ensure the purity of the samples, we adopt a false-alarm rate (FAR) of $0.25 {\rm yr}^{-1}$ as the threshold to select the events.
We exclude GW190814 in the analysis because the low secondary mass ($\sim2.6M_{\odot}$) makes it disconnected from the BBH population \citep{2021ApJ...913L...7A,2022ApJ...926...34E} but potentially connected to the recently identified population of NSBHs \citep{2021ApJ...907L..24S,2021ApJ...922....3T}. 
Therefore, we adopt a sample of 62 BBHs based on our criteria of FAR $< 0.25 {\rm yr}^{-1}$.
The posterior samples for each BBH event are adopted from Gravitational Wave Open Science Center (\url{https://www.gw-openscience.org/eventapi/html/GWTC/}).
For the (new) events in the GWTC-1 \citep{2019PhRvX...9c1040A}, GWTC-2 \citep{2021PhRvX..11b1053A}, GWTC-2.1 \citep{2021arXiv210801045T} and GWTC-3 \citep{2021arXiv211103606T}, we use the `Overall posterior' samples, 
the `PublicationSamples' samples,
the `PrecessingSpinIMRHM' samples, 
and the `C01:Mixed' samples, respectively.

\begin{figure*}
	\centering  
	\subfigbottomskip=2pt 
	\subfigcapskip=-5pt 
	\subfigure[]{
		\includegraphics[width=0.48\linewidth]{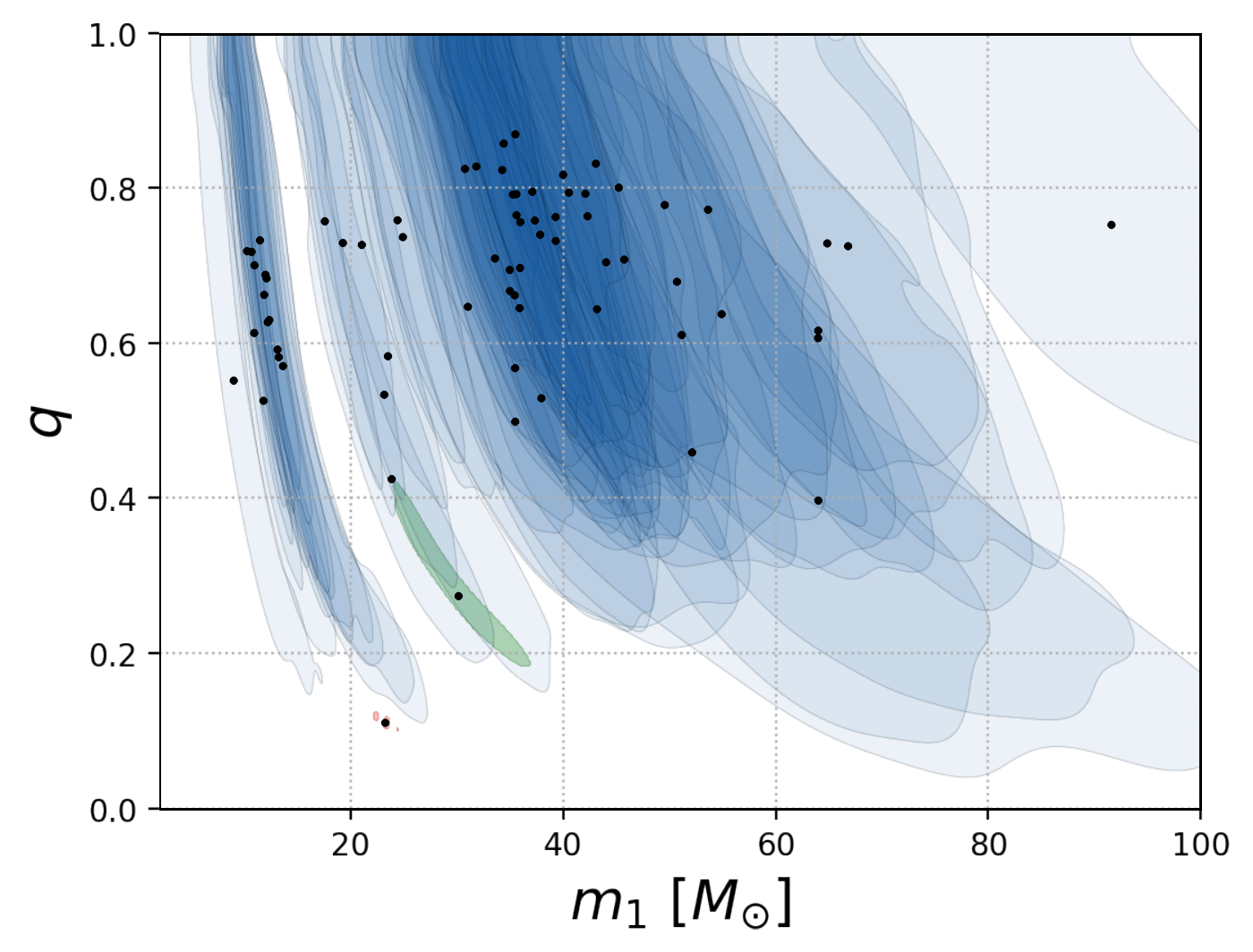}}
	\subfigure[]{
		\includegraphics[width=0.48\linewidth]{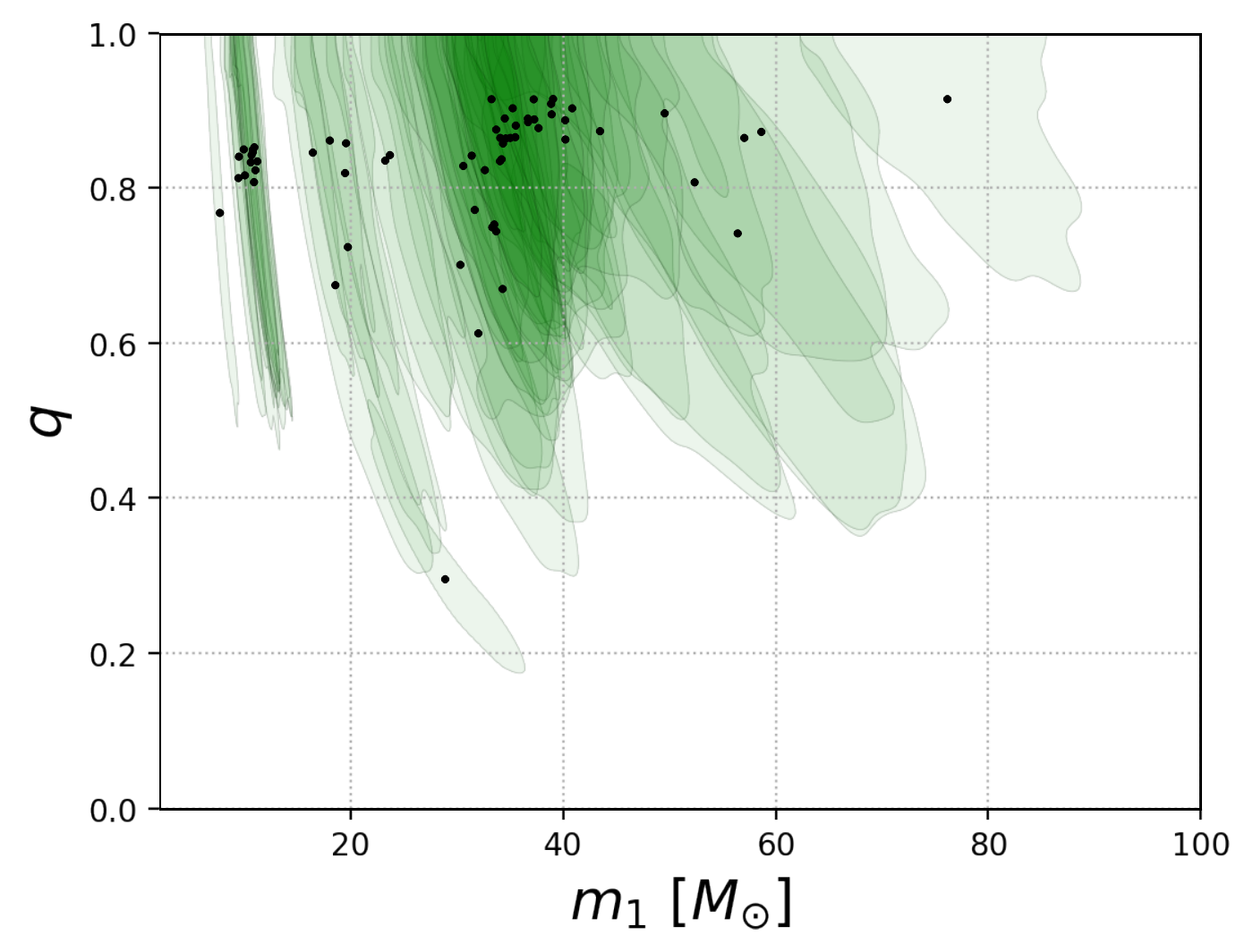}}
	\subfigure[]{
		\includegraphics[width=0.48\linewidth]{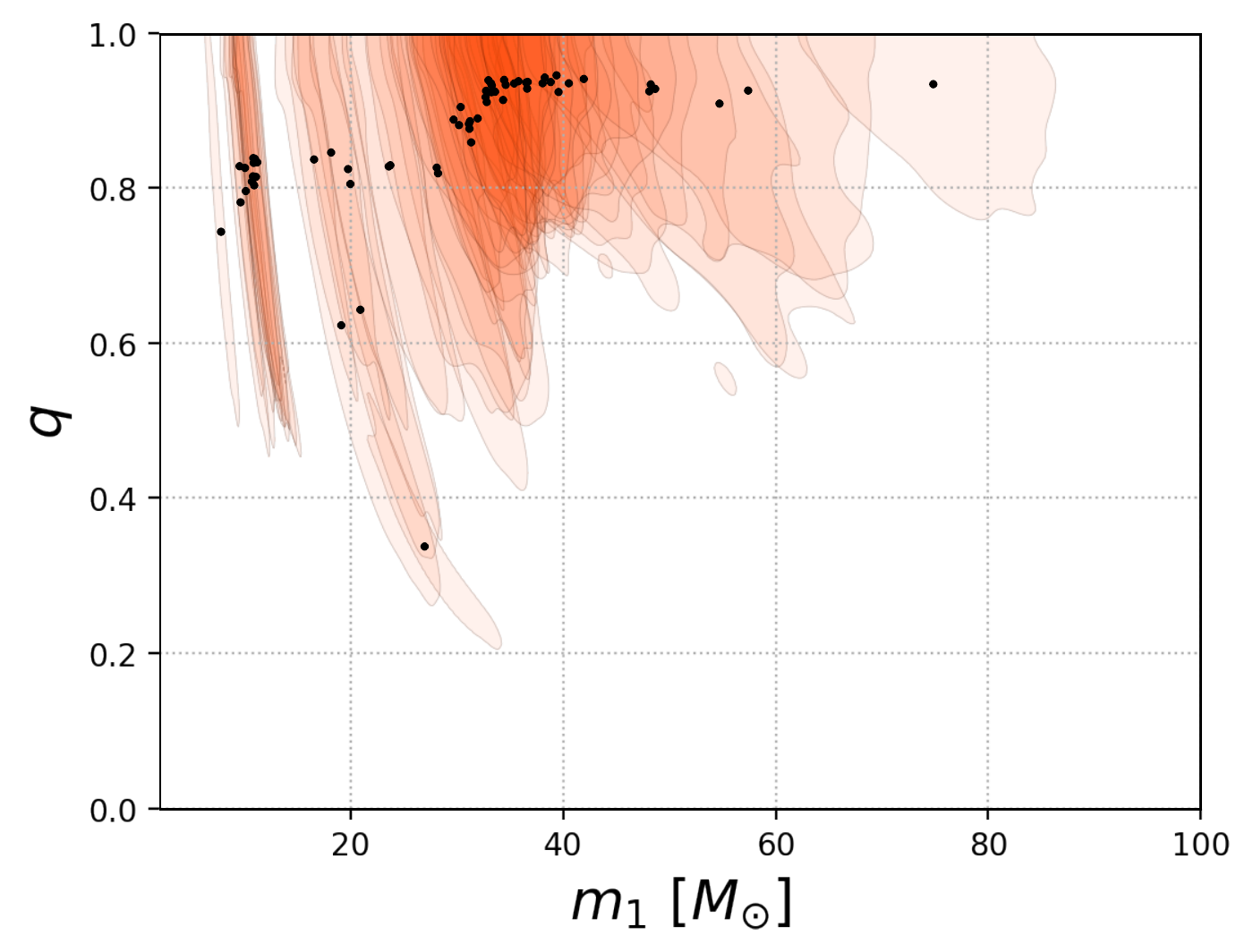}}
	\subfigure[]{
		\includegraphics[width=0.48\linewidth]{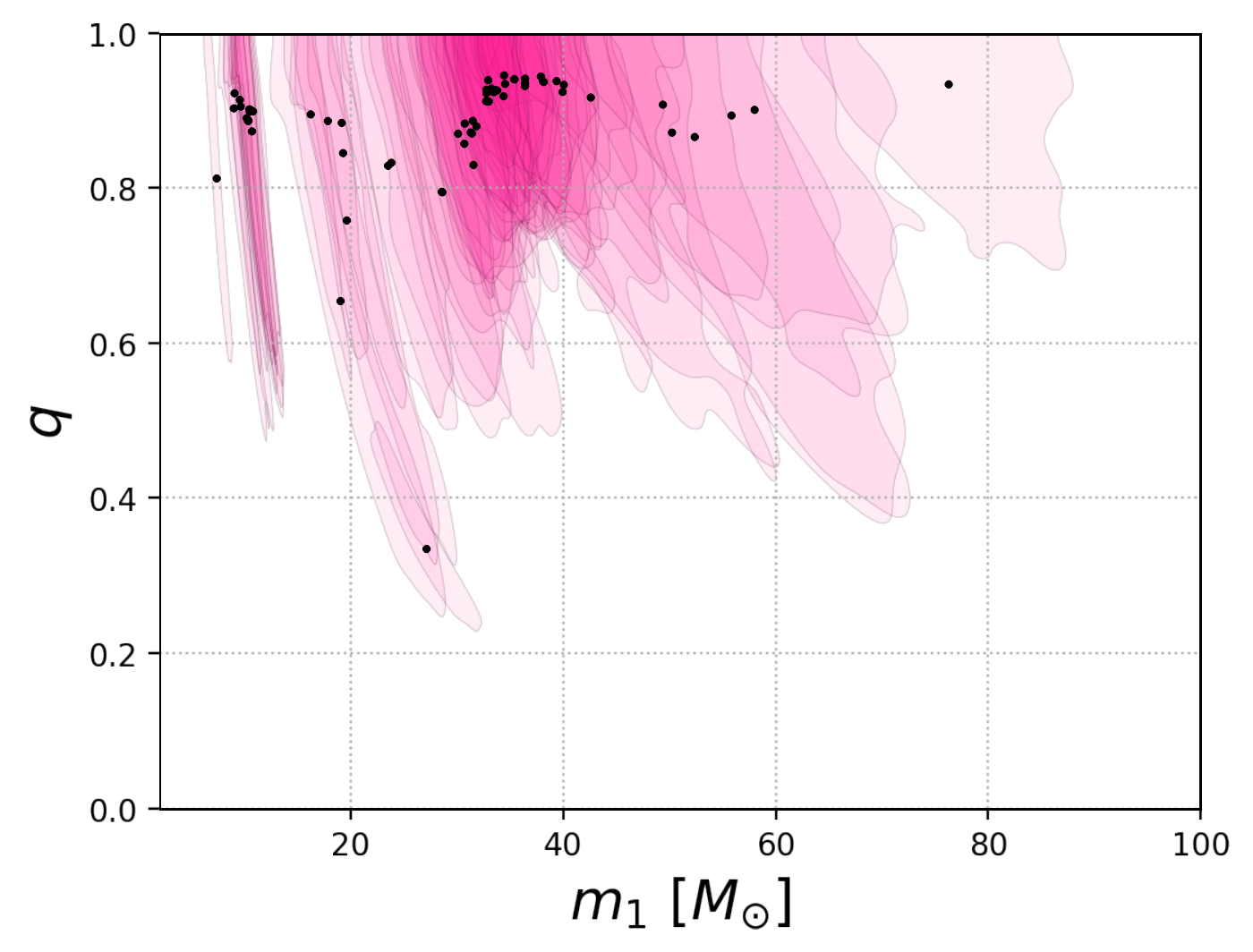}}
\caption{(a) Primary mass and mass ratio posteriors for the 63 BBH candidates (including GW190814) in the LVKC GWTC-3 catalog \citep{2021arXiv211103606T} with FARs below $0.25 {\rm yr}^{-1}$, as obtained under a default prior. Each shaded region (black point) represents the central 90\% credible posterior bounds (median $m_1$ and median $q$ value) of a given BBH. The red and green regions represent the three asymmetric systems GW190814 and GW190412.
(b) The same as (a), but for the posterior samples are reweighed by \textsc{PowerLaw Peak} with the half-Gaussian mass ratio model (Eq. (\ref{sigma_q})) described in Sec. \ref{method_model}.
(c) The same as (a), but for the posterior samples are reweighed by \textsc{PowerLaw Peak} with the $m_1$-dependent mass ratio model of Eq. (\ref{model1}) described in Sec. \ref{method_model}.
(d) The same as (a), but for the posterior samples are reweighed by \textsc{PowerLaw Peak} with the $m_1$-dependent mass ratio model of Eq. (\ref{model2}) described in Sec. \ref{method_model}.}
\label{post_events}
\end{figure*}

\subsection{Models} \label{method_model}
With the updated data from the GWTC-3, the simple \textsc{PowerLaw  Peak} model is still acceptable \citep{2021arXiv211103634T}. 
Therefore, in our analysis, the distribution of the primary masses is described by
\begin{equation*}\label{PLP}
\begin{aligned}
\pi(m_1&| \alpha, m_{\rm min}, m_{\rm max}, \delta_{\rm m}, \lambda, \mu, \sigma) \\
&= ((1-\lambda)\mathcal{P}(m_1 | -\alpha, m_{\rm min}, m_{\rm max}, \delta_{\rm m})\\
&+\lambda\mathcal{G}(m_1 | \mu, \sigma, m_{\rm min}, m_{\rm max})),
\end{aligned}
\end{equation*}
where $m_1$ is the primary mass of the BBHs; $-\alpha$, $\delta_{\rm m}$, are the power-law spectral index and smoothing scale; $\lambda$, $\mu$, and $\sigma$ are the mixing fraction, mean, and width of the \textsc{Peak} component; $m_{\rm min}$ and $m_{\rm max}$ are the low-mass and high-mass cutoff.
Note that the power-law component $\mathcal{P}$ is normalized after the smoothing treatment on its lower boundary as suggested by \citet{2021ApJ...913...42W}, i.e., 
\begin{equation*}
\begin{aligned}
\mathcal{P}(m &| -\alpha, m_{\rm min}, m_{\rm max}, \delta_{\rm m}) \\
&= A_1m^{-\alpha}S(m|m_{\rm min}, \delta_{\rm m}), ~ \text{for}~ m \in{(m_{\rm min}, m_{\rm max})},
\end{aligned}
\end{equation*}
where $A_1$ is the normalization constant and $S$ is the smoothing function (see \cite{2021ApJ...913L...7A} and \cite{2021arXiv211103634T} for details).

In this work, we also use the \textsc{Broken PowerLaw} \citep[see][for detail]{2021ApJ...913L...7A} to model the primary mass distribution for a checking purpose,
\begin{equation}\label{BPL}
\begin{aligned}
\pi(m_1&| \alpha_1, m_{\rm min}, m_{\rm max}, \delta_{\rm m}, b, \alpha_2)\\
 \propto  &
\begin{cases}
\begin{aligned}
m_1^{-\alpha_1}&S(m_1|m_{\rm min},  \delta_{\rm m}),\\ &\text{for}~  m_{\rm min}< m_1 <m_{\rm min}+b(m_{\rm max}-m_{\rm min}), \\
m_1^{-\alpha_2}&S(m_1|m_{\rm min},  \delta_{\rm m}),\\ &\text{for}~  m_{\rm min}+b(m_{\rm max}-m_{\rm min})<  m_1 < m_{\rm max}, \\
 0,~~~~ &~~~~~~~~~~~~~~\text{otherwise.}
 \end{aligned}
\end{cases}
\end{aligned}
\end{equation}
where $-\alpha_1$ and $-\alpha_2$ are the power-law slopes in the low- and high-mass ranges, and $b$ is the fraction of the way between $m_{\rm min}$ and $m_{\rm max}$ at which the primary mass distribution breaks.

The distribution of the secondary masses $\pi(m_2| m_1, \boldsymbol{\Lambda_2})$, with respect to $m_1$, is potentially associated with the pairing function, where $\boldsymbol{\Lambda_2}$ is the hyperparameters for the $m_2$ distribution. A simple description for $m_2$ distribution is the power-law model widely adopted in the literature \citep[e.g.,][]{2021ApJ...913L...7A,2021arXiv211103634T}
\begin{equation}\label{beta}
\begin{aligned}
\pi(&m_2| m_1, \beta, \delta_{\rm m})\\
& = A_2m_2^{\beta}S(m_2|m_{\rm min}, \delta_{\rm m}), ~ \text{for}~ m_2 \in{(m_{\rm min}, m_1)},
\end{aligned}
\end{equation}
where $A_2$ is the normalization constant. In this model, the constraint on the  power-law slope $\beta$ of this model is found to be sensitive to whether some highly asymmetric events like GW190412\footnote{Note that GW190412 is not an outlier because the posterior of $\beta$ inferred from the inclusion of GW190412 has a significant overlap with the leave-one-out posterior \citep[see][for details]{2021ApJ...913L...7A}.} are included in the hierarchical analysis \citep[see][for details]{2020PhRvD.102d3015A}. Therefore in this work, we apply another simple parameterization for $m_2$ distribution,
\begin{equation}\label{eq_sigma_q}
\begin{aligned}
\pi(&m_2| m_1, \sigma_{\rm q})\\
& = A_3\mathcal{G}(m_2 | m_1,m_1\sigma_{\rm q}) ~ \text{for}~ m_2 \in{(m_{\rm min}, m_1)},
\end{aligned}
\end{equation}
i.e., a half-Gaussian with peak value and width of $m_1$ and $\sigma_{\rm q}m_1$, respectively, where $A_3$ is the normalization constant.

As shown in the posteriors obtained by the default prior (see Fig. \ref{post_events} (a)), many BBHs with primary masses in $\sim (30,40)M_{\odot}$ (i.e, the location of the \textsc{Peak} in the $m_1$ distribution) show a stronger tendency to have equal masses, though most of the other BBHs are consistent with symmetric systems. When the posteriors are reweighed by a population model with a secondary mass distribution of Eq. (\ref{eq_sigma_q}), such a feature becomes more obvious as shown in Fig. \ref{post_events} (b). 
To examine whether and how the mass ratio distribution varies in the whole mass range, we introduce some $m_1$-dependent mass ratio distribution models. These models take Eq. (\ref{eq_sigma_q}) as a prototype, but with a $\sigma_{\rm q}$ varying with $m_1$. 
Firstly, we divide the BBHs into two mass ranges ($m_1<m_{\rm cut}$ and $m_1>m_{\rm cut}$), where the BBHs have two different mass ratio distributions, hereafter Model \uppercase\expandafter{\romannumeral1}, 
\begin{equation}\label{model1}
\begin{aligned}
\sigma_{\rm q}(m_1 &| \sigma_{\rm q}^{\rm low}, \sigma_{\rm q}^{\rm high},m_{\rm cut})\\&=
\begin{cases}
\sigma_{\rm q}^{\rm low},&\text{for}~  m_1<m_{\rm cut}, \\
\sigma_{\rm q}^{\rm high},&\text{for}~  m_1>m_{\rm cut}.
\end{cases}
\end{aligned}
\end{equation}
To further check how the mass ratio distribution varies with the primary mass in detail, we use the cubic spline to interpolate the $\sigma_{\rm q}(x)$ that describes the mass ratio distribution of BBHs with $m_1=x$; such a nonparametric method was initially used to characterize the primary mass function by \cite{2022ApJ...924..101E}. So the corresponding formula is (hereafter Model \uppercase\expandafter{\romannumeral2}), 
\begin{equation}\label{model2}
\begin{aligned}
\sigma_{\rm q}(m_1& | \sigma_{\rm q}^{\rm left}, \sigma_{\rm q}^{\rm right}, m_{\rm min}, m_{\rm max},\{m_i,f_i\}_{i=1}^{N_{\rm knot}})\\
& = \frac{(m_1-m_{\rm min})\sigma_{\rm q}^{\rm right}+(m_{\rm max}-m_1)\sigma_{\rm q}^{\rm left}}{m_{\rm max}-m_{\rm min}}\\
& \times \text{exp}(f(m_1;\{m_i,f_i\}_{i=1}^{N_{\rm knot}}))
\end{aligned}
\end{equation}
where $f(m_1;\{m_i,f_i\}_{i=1}^{N_{\rm knot}})$ is the perturbation function modeled as a cubic spline that is interpolated between $N_{\rm knot}$ knots placed in $m_1$ space, and its shape can be determined by the heights $\{f_i\}$ at their knots $\{m_i\}$. Following \cite{2022ApJ...924..101E}, we fix the locations of each knot to be linear in log $m_1$ space of $(5,100)M_{\odot}$, and restrict the perturbation to zero at the minimum and maximum knots. We find that 15 knots can make it flexible enough to characterize the mass ratio distribution in the whole primary mass range in detail.

We also try to find out whether the mass ratio distributions are different in the two components of the $m_1$ distribution (i.e., the \textsc{PowerLaw} and the \textsc{Peak}). The distributions of the secondary masses in the two components are described by Eq. (\ref{eq_sigma_q}) with $\sigma_{\rm q}^{\rm PL}$ and $\sigma_{\rm q}^{\rm G}$. So, the Model \uppercase\expandafter{\romannumeral3} reads 
\begin{equation}
\begin{aligned}
\pi(m_1&| \alpha, m_{\rm min}, m_{\rm max}, \delta_{\rm m}, \lambda, \mu, \sigma, \sigma_{\rm q}^{\rm PL}, \sigma_{\rm q}^{\rm G}) 	\\
&= (1-\lambda)\mathcal{P}(m_1 | -\alpha, m_{\rm min}, m_{\rm max}, \delta_{\rm m})\pi(m_2| m_1, \sigma_{\rm q}^{\rm PL}) \\
&+\lambda\mathcal{G}(m_1 | \mu, \sigma, m_{\rm min}, m_{\rm max})\pi(m_2| m_1, \sigma_{\rm q}^{\rm G}) .
\end{aligned}
\end{equation}

All the parameters, their descriptions, and the priors are summarized in Tab. \ref{prior}. Because there is a mass-spin degeneracy, we fit the distribution of primary and secondary masses jointly with the spin distribution and the \textsc{Default} spin model as defined in \cite{2021arXiv211103634T} is adopted. 

\subsection{Hierarchical inference}
We perform a hierarchical Bayesian inference to fit the data of the observed events \{$d$\} with the population models described above. Following the framework described in \cite{2021ApJ...913L...7A} and \cite{2021arXiv211103634T}, for the given data $\{d\}$ from $N_{\rm det}$ GW detections, the likelihood of the hyperparameters $\boldsymbol{\Lambda}$ can be expressed as 
\begin{equation}\label{eq_llh}
\mathcal{L}(\{d\} |\boldsymbol{\Lambda})\propto N^{N_{\rm det}}e^{-N{\xi(\boldsymbol{\Lambda})}}\prod_{i=1}^{N_{\rm det}}\int{\mathcal{L}(d_i|\theta_i)\pi(\theta_i|\boldsymbol{\Lambda})d\theta_i},
\end{equation} 
where $N$ is the number of mergers in the universe over the observation period, which is related to the merger rate, and $\xi(\boldsymbol{\Lambda})$ means the detection fraction.
The single-event likelihood $\mathcal{L}(d_i|\theta_i)$ can be estimated using the posterior samples (see \cite{2021ApJ...913L...7A} for detail), and $\xi(\boldsymbol{\Lambda})$ is estimated using a Monte Carlo integral over detected injections as introduced in the Appendix of \cite{2021ApJ...913L...7A}. We assume that the merger rate density increases with redshift, $\mathcal{R}\propto(1 + z)^{2.7}$ as obtained by \cite{2021arXiv211103634T}.
The injection campaigns can be adopted from \cite{ligo_scientific_collaboration_and_virgo_2021_5546676}, where they combine the O1, O2, and O3 injection sets, ensuring a constant rate of injections across the total observing time.
We apply the sampler \textit{Pymultinest} \citep{2016ascl.soft06005B} for the hierarchical Bayesian inference of the posteriors.

\section{Results}\label{sec:result}

\begin{table*}[htpb]\label{prior}
\begin{ruledtabular}
\caption{Hyperparameters, Their Descriptions, and Chosen Priors for This Work for Each Respective Population Model}
\begin{tabular}{cccc}
Models   & parameters & descriptions & priors \\
\cline{1-4}
\multicolumn{4}{c}{Primary mass distribution models}  \\ 
\cline{1-4}
\multirow{7}{*}{\textsc{PowerLaw Peak}} & $\alpha$ & slope of the power law & U(-4,12) \\
&$m_{\rm min}$ &minimum mass cutoff&U(2,10)\\
&$m_{\rm max}$&maximum mass cutoff&U(50,100)\\
&$\delta_{\rm m}$& width of mass range that smoothing function impact on&U(0,10) \\
&$\mu$ & center of the Gaussian component &  U(20,50)\\
&$\sigma$ & width of the Gaussian component  & U(0.5,10)\\
&$\lambda$ & fraction of BBH in the Gaussian component & U(0,1)\\
\cline{1-4}
\multirow{6}{*}{\textsc{Broken PowerLaw}} & $\alpha_1$ & slope of the first power law & U(-4,12) \\
& $\alpha_2$ & slope of the second power law & U(-4,12) \\
&$m_{\rm min}$ &minimum mass cutoff&U(2,10)\\
&$m_{\rm max}$&maximum mass cutoff&U(50,100)\\
&$\delta_{\rm m}$&width of mass range that smoothing function impact on&U(0,10) \\
&$b$& fraction between $m_{\rm min}$ and $m_{\rm max}$ where the power law break lies & U(0,1) \\
\cline{1-4}
\multicolumn{4}{c}{Mass ratio distribution models}  \\ 
\cline{1-4}
\multirow{1}{*}{Half-Gaussian} & $\log_{10}\sigma_{\rm q}$ & logarithmic width of the mass ratio distribution & U(-2,0) \\
\cline{1-4}
\multirow{3}{*}{Model \uppercase\expandafter{\romannumeral1}} & $\log_{10}\sigma_{\rm q}^{\rm low}$ & $\log_{10}\sigma_{\rm q}$ in the lower-mass range& U(-2,0) \\
& $\log_{10}\sigma_{\rm q}^{\rm high}$ & $\log_{10}\sigma_{\rm q}$ in the higher-mass range& U(-2,0) \\
& $m_{\rm cut}$ & point dividing the lower and higher-mass ranges & U(20,40) \\
\cline{1-4}
\multirow{3}{*}{Model \uppercase\expandafter{\romannumeral2}} & $\log_{10}\sigma_{\rm q}^{\rm left}$ & $\log_{10}\sigma_{\rm q}$ at the lower-mass  edge& U(-2,0) \\
& $\log_{10}\sigma_{\rm q}^{\rm right}$ & $\log_{10}\sigma_{\rm q}$ at the higher-mass  edge& U(-2,0) \\
& $\{f_i\}_{i=1}^{15}$ & y-value of the spline interpolant knots & $\mathcal{N}(0, \sigma_{\rm knot})$\\
\cline{1-4}
\multirow{2}{*}{Model \uppercase\expandafter{\romannumeral3}} & $\log_{10}\sigma_{\rm q}^{\rm PL}$ & $\log_{10}\sigma_{\rm q}$ in the power-law component& U(-2,0) \\
& $\log_{10}\sigma_{\rm q}^{\rm G}$ & $\log_{10}\sigma_{\rm q}$ in the Gaussian component& U(-2,0) \\
\end{tabular}
\tablenotetext{}{{\bf Note.} Here, `U' means the uniform distribution.}
\end{ruledtabular}
\end{table*}

\begin{figure*}
\centering  
	\subfigbottomskip=1pt 
	\subfigcapskip=-1pt 
	\subfigure[]{
		\includegraphics[width=0.49\linewidth]{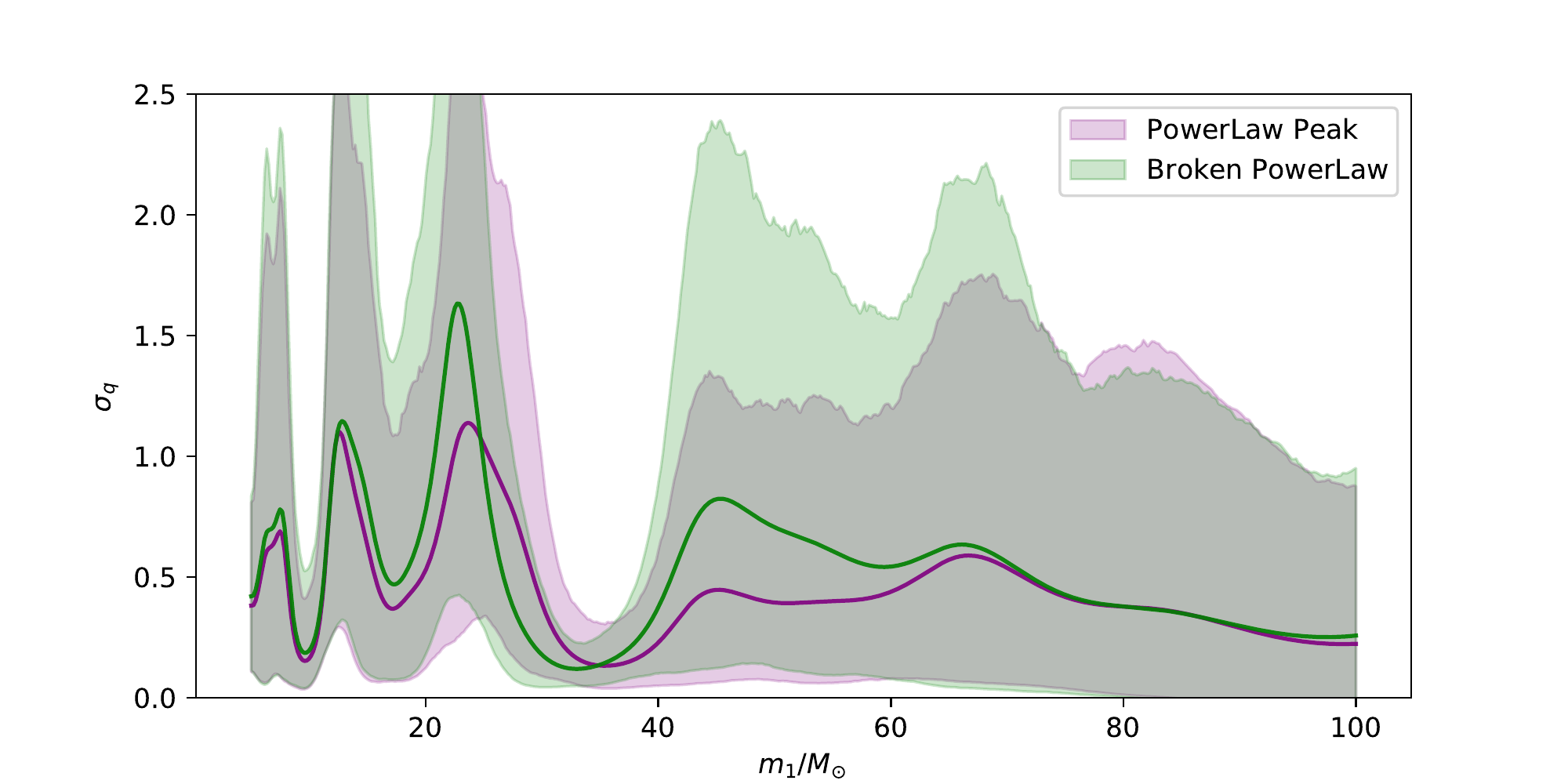}}
	\subfigure[]{
		\includegraphics[width=0.49\linewidth]{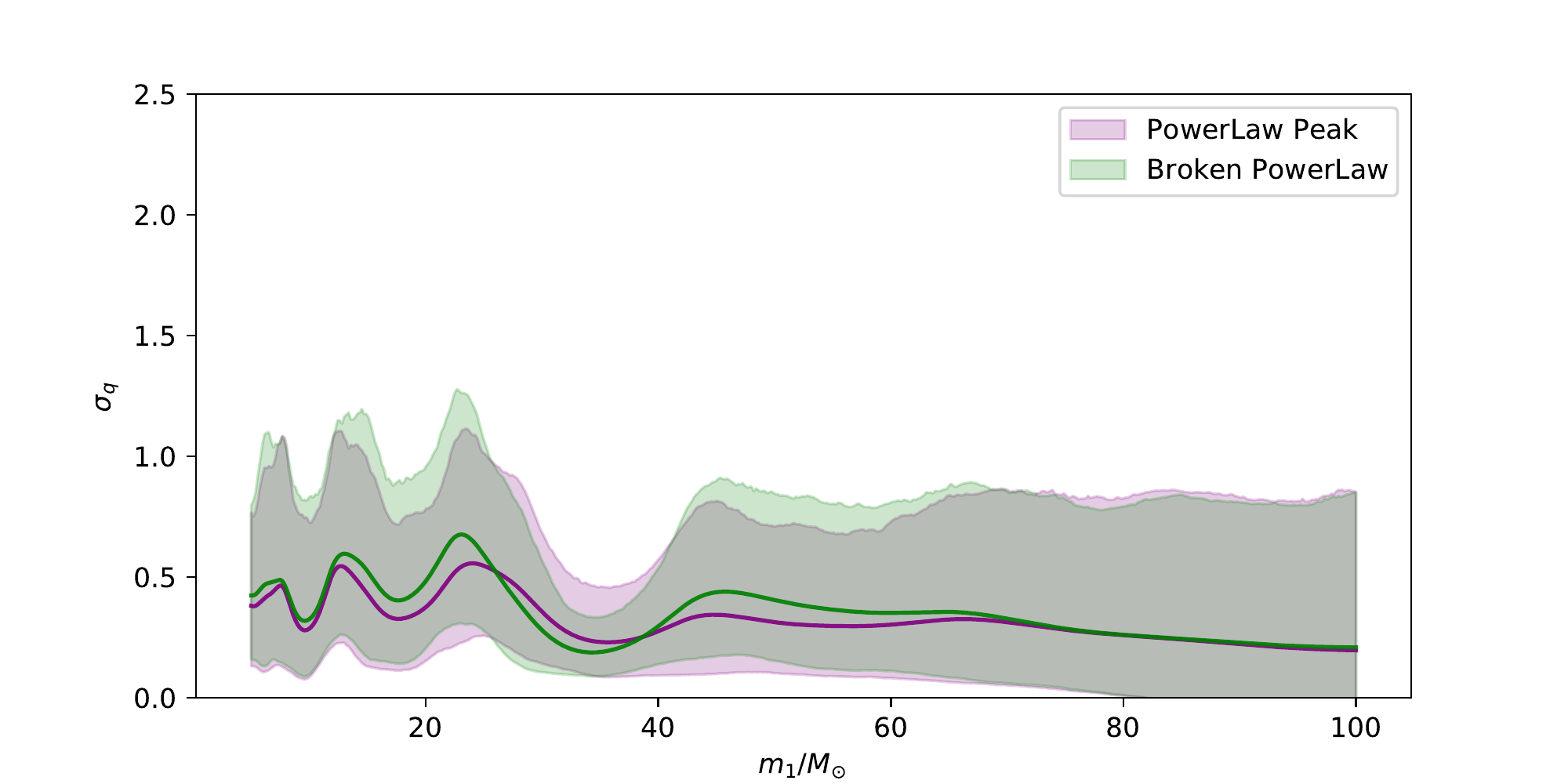}}
\caption{1 $\sigma$ width of the mass ratio distribution as a function of the primary mass; the shaded regions stand for the 90\% credible interval, and the solid curves are the mean values. (a) and (b) are the results that obtained using Model \uppercase\expandafter{\romannumeral2} with $\sigma_{\rm knot}=1.0$ and $\sigma_{\rm knot}=0.5$. }
\label{sigma_q}
\end{figure*}

In this section, we display the results obtained using the methods described in Section \ref{sec:method}. All the results shown here are marginalized over the hyperparameters of the spin distribution. We firstly compare all the models by the Bayes factors, as summarized in Table \ref{BF}. 
Model \uppercase\expandafter{\romannumeral2} provides us with an overall picture \footnote{Note that the lower bound of $\sigma_{\rm q}$ in the higher-mass range can still be smaller than that in the range of 30-40 $M_{\odot}$.} of how the mass ratio distribution varies with the primary mass, as is shown in Fig. \ref{sigma_q}. 
We find nontrivial Bayes factors between Model \uppercase\expandafter{\romannumeral2} (with $\sigma_{\rm knot}$=0.5 or $\sigma_{\rm knot}$=1 ) and invariable half-Gaussianmodel.
Additionally, the Bayes factors are still significantly positive when we change the primary mass model or exclude the asymmetric event GW190412. 
So, we conclude that an invariable mass ratio distribution in the whole mass range is disfavored. The mass ratio distribution should vary with primary mass, which indicates that the BBHs in the different mass ranges may have different evolutionary processes.

\begin{figure*}
	\centering  
	\subfigbottomskip=2pt 
	\subfigcapskip=-5pt 
	\subfigure[]{
		\includegraphics[width=0.48\linewidth]{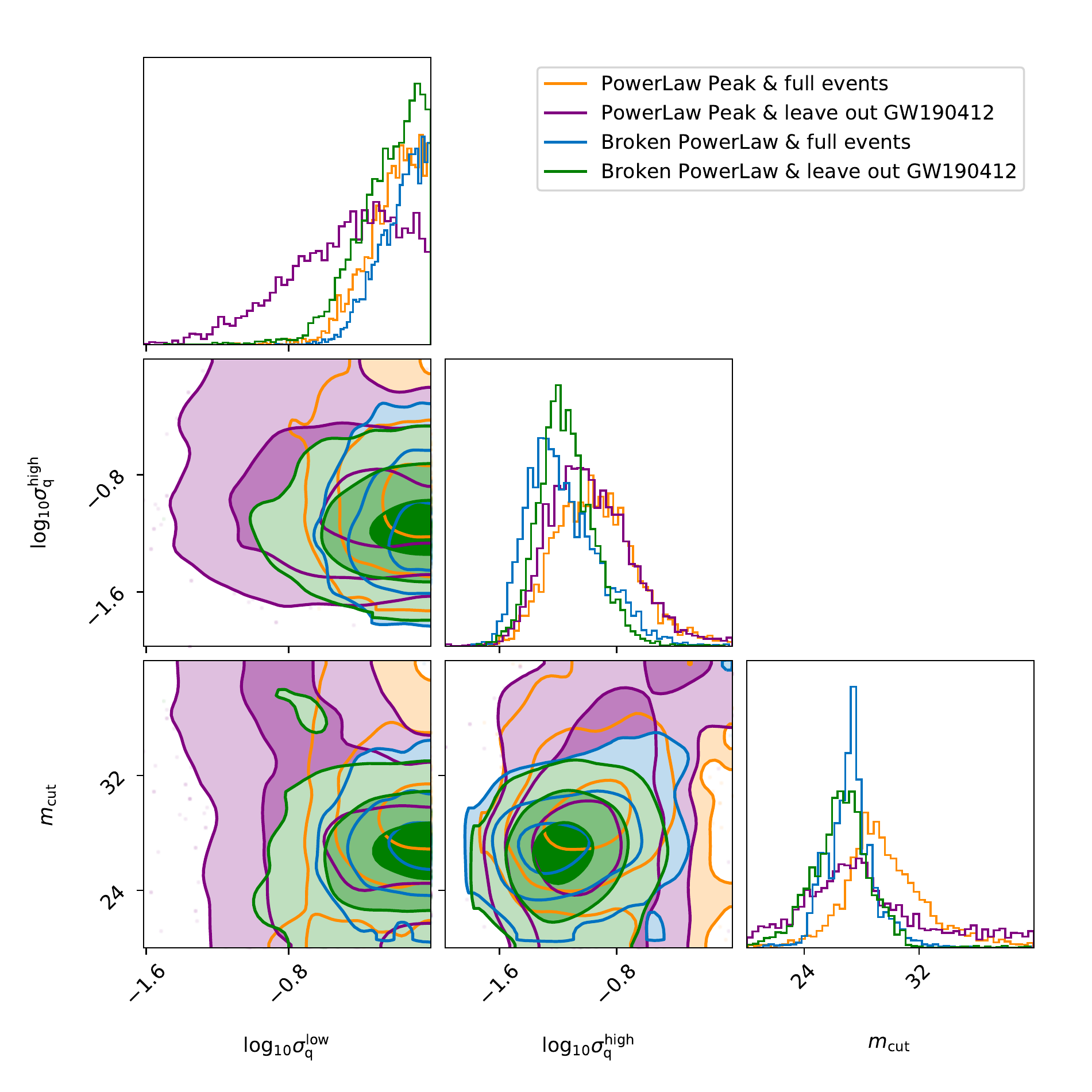}}
	\subfigure[]{
		\includegraphics[width=0.48\linewidth]{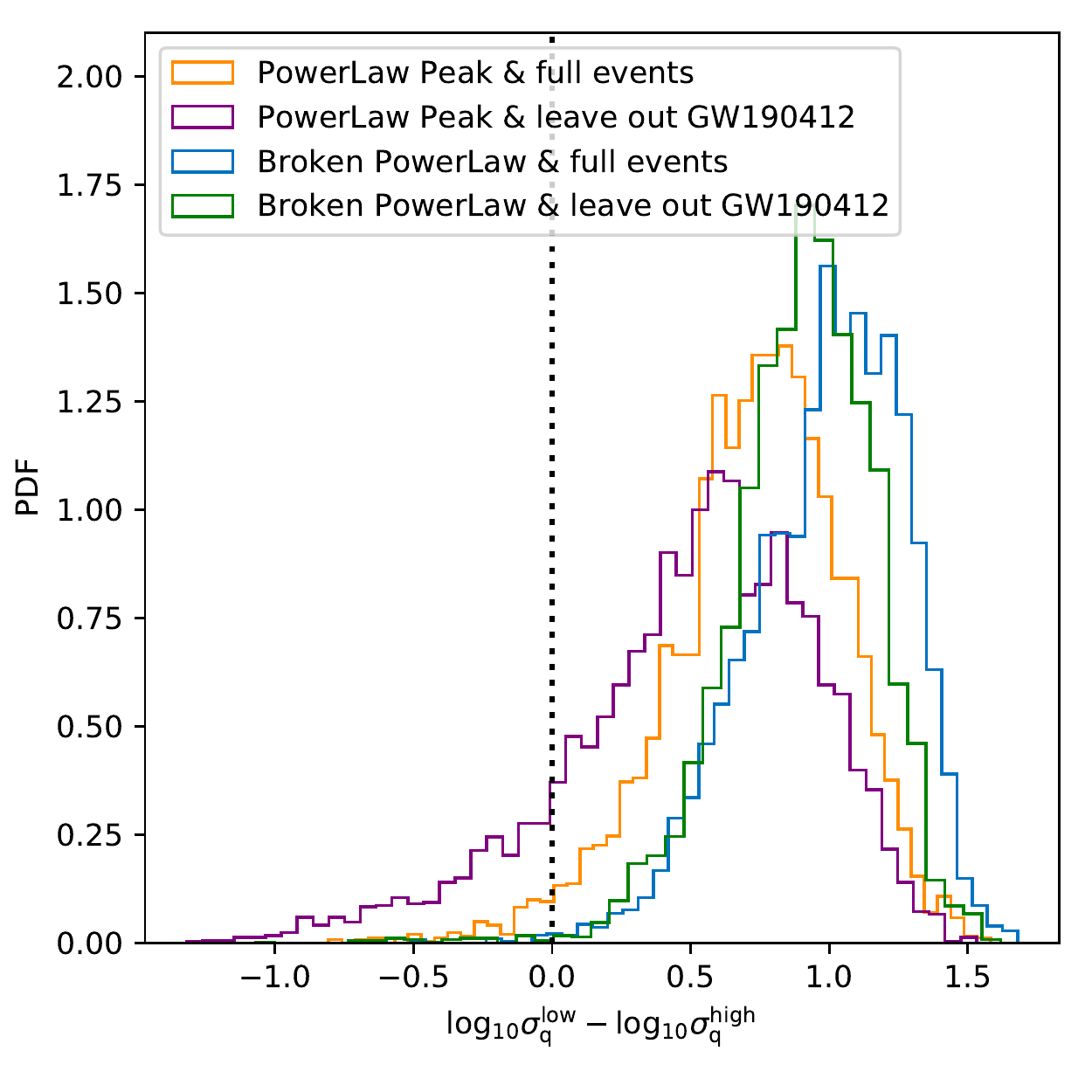}}
\caption{(a) Posterior distributions of $\log_{10}\sigma_{\rm q}^{\rm low}$ ($\log_{10}\sigma_{\rm q}^{\rm high}$) that describe the mass ratio distribution of BBHs in the lower (higher) mass ranges, and the split point $m_{\rm cut}$ obtained by Model\uppercase\expandafter{\romannumeral1}. (b) Posterior distributions of $\log_{10}\sigma_{\rm q}^{\rm low}-\log_{10}\sigma_{\rm q}^{\rm high}$, when we change the primary mass distribution model, the value is still confidently positive, but the credibility becomes lower in the absence of GW190412.}
\label{mcut}
\end{figure*}

Model \uppercase\expandafter{\romannumeral1} divides the BBHs into two mass ranges (low and high) at $m_{\rm cut}$, where the mass ratio distributions have two different widths, $\sigma_{\rm q}^{\rm low}$ and $\sigma_{\rm q}^{\rm high}$. 
We obtained $m_{\rm cut}=29.3^{+5.7}_{-4.0}M_{\odot}$ (90\% credible interval), and the value slightly shifts to a smaller value of $27.4^{+10.2}_{-5.5}M_{\odot}$ if we leave out GW190412 from the analysis.
We find that the BBHs in the high-mass range (i.e., with $m_1>m_{\rm cut}$) have a more preference for equal masses than those in the low-mass range (i.e., with $m_1< m_{\rm cut}$) at a $97.6\%$ credible level, as shown in Fig. \ref{mcut} (a), and the credibility becomes $86.7\%$ in the absence of GW190412.
To find out the influence of the primary mass model, we also perform the inferences with the \textsc{Broken PowerLaw} model \citep{2021ApJ...913L...7A} instead of the \textsc{PowerLaw Peak} (Eq. (\ref{PLP})); then the conclusion remains unchanged, and the credibility even rises to $99.7\%$, though $m_{\rm cut}$ slightly shift to lower values, as shown in Fig. \ref{mcut} (b).
As shown in Fig. \ref{sigma_q} that $\sigma_{\rm q}$ may wiggle in the lower-mass range ($m_1<m_{\rm cut}$). To find out whether there are actually additional features, We use an extended Model \uppercase\expandafter{\romannumeral1} (see the Appendix \ref{app_lowmass}) for analysis. Our results suggest that none of the perturbations are statistically significant enough to declare an additional structure as shown in Fig. \ref{model4_fm}.

\begin{figure*}
	\centering  
	\subfigbottomskip=2pt 
	\subfigcapskip=-5pt 
	\subfigure[]{
		\includegraphics[width=0.48\linewidth]{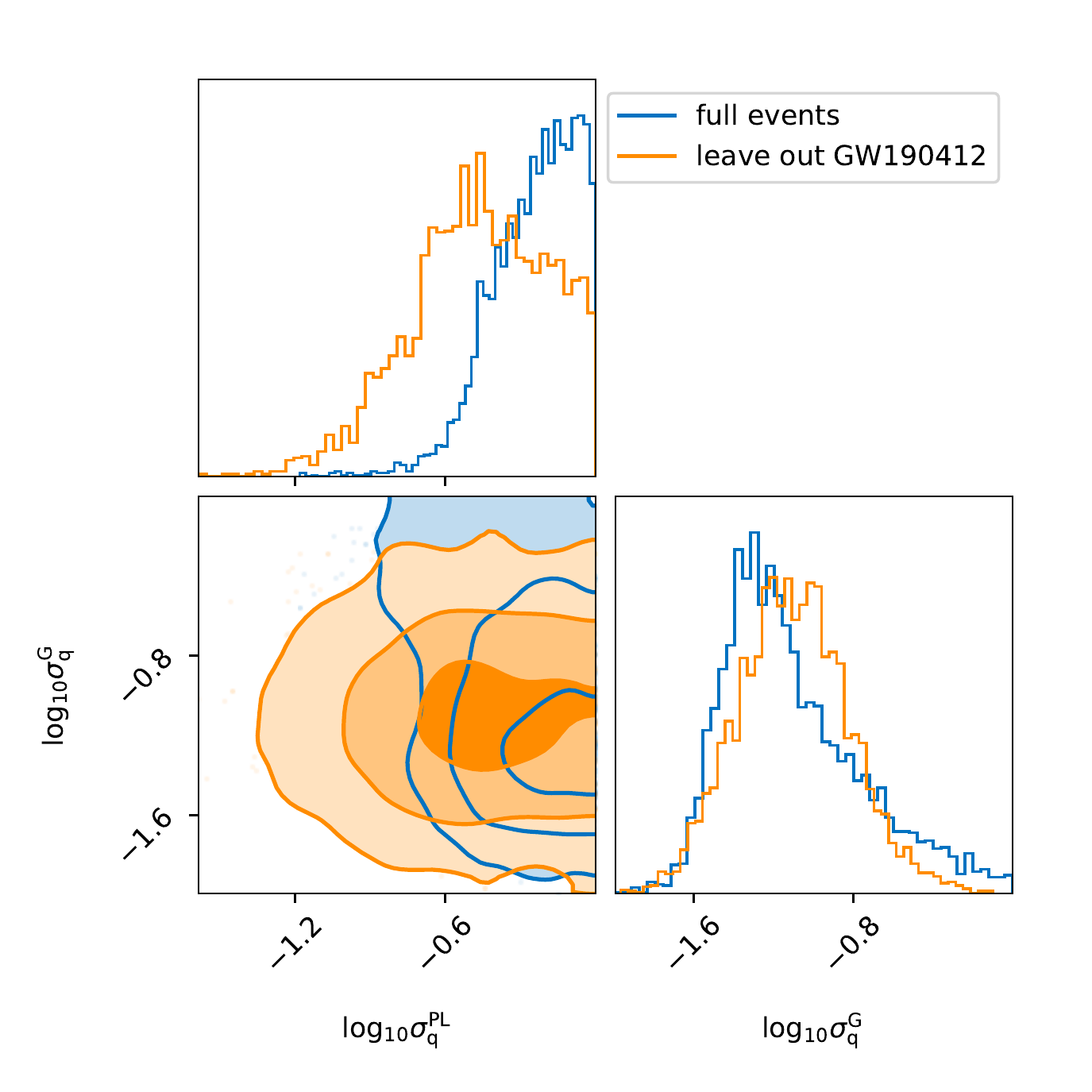}}
	\subfigure[]{
		\includegraphics[width=0.48\linewidth]{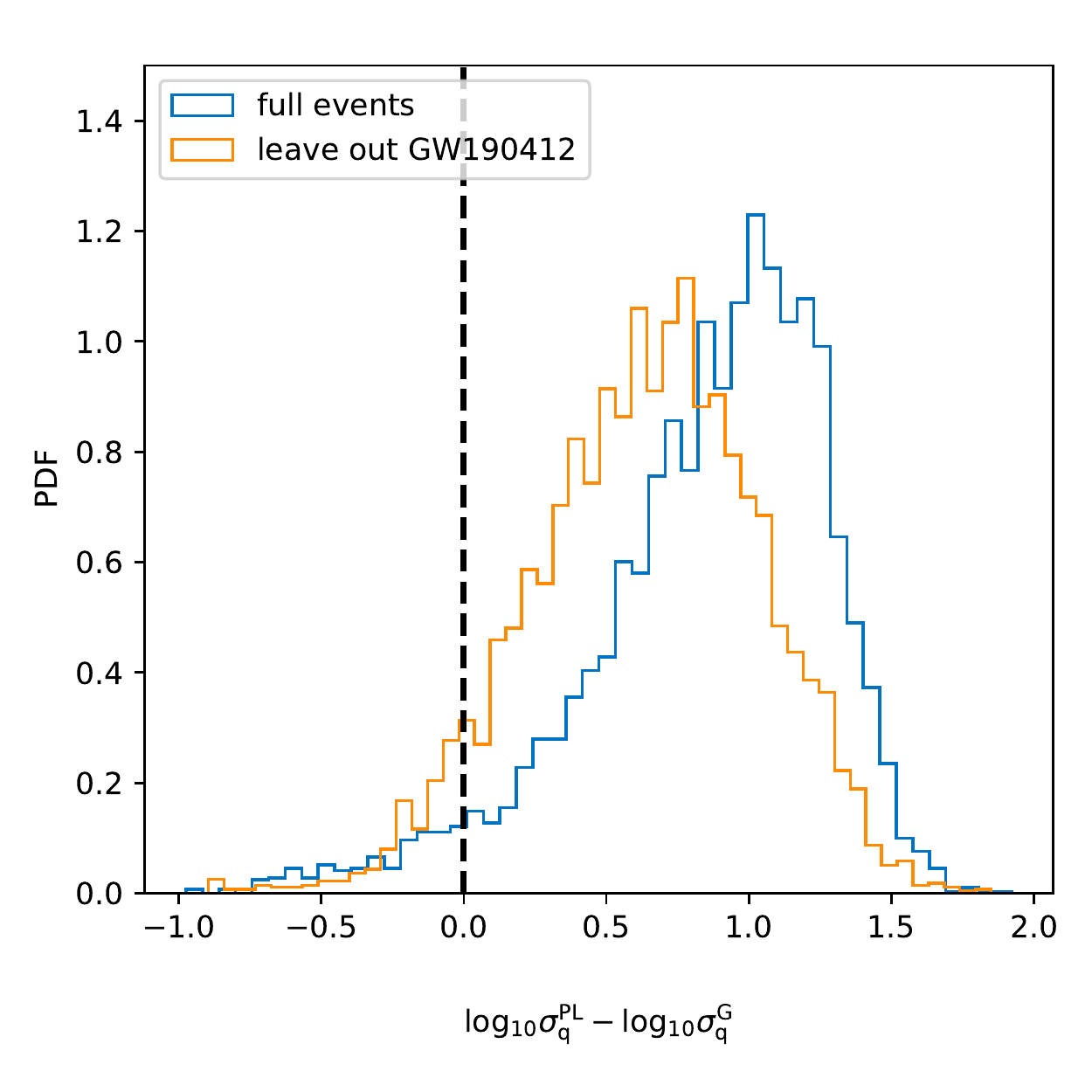}}
\caption{(a): Posterior distributions of $\log_{10}\sigma_{\rm q}^{\rm PL}$ ($\log_{10}\sigma_{\rm q}^{\rm G}$) that describe the mass ratio distribution of BBHs in the {\bf Power Law (Gaussian) component }obtained by Model\uppercase\expandafter{\romannumeral3}; (b): Posterior distributions of $\log_{10}\sigma_{\rm q}^{\rm PL}-\log_{10}\sigma_{\rm q}^{\rm G}$. When we exclude the GW190412, $\log_{10}\sigma_{\rm q}^{\rm PL}$ shifts to a smaller value, but the value of $\log_{10}\sigma_{\rm q}^{\rm PL}-\log_{10}\sigma_{\rm q}^{\rm G}$ is still confidently positive.}
\label{find_twins}
\end{figure*}

The Bayes factors between Model \uppercase\expandafter{\romannumeral3} and  the invariable half-Gaussian mass ratio distribution model as summarized in Tab. \ref{BF}, provide milder evidence that the BBHs with $m_1$ in the Gaussian component have a mass ratio distribution different from that of the other mass range, as shown in Fig. \ref{find_twins} (a). The width of the mass ratio distribution of BBHs in the Gaussian component ($\sigma_{\rm q}^{\rm G}$) is smaller than that in the power-law component ($\sigma_{\rm q}^{\rm PL}$) at $95.4\%$ credibility (see Fig. \ref{find_twins} (b));
such a feature is also present in Fig. \ref{sigma_q} as well as Fig. \ref{post_events}, where the BBHs with $m_1\sim35M_{\odot}$ have a stronger preference for being equal-mass systems. 
When we exclude GW190412 from the analysis, $\sigma_{\rm q}^{\rm PL}-\sigma_{\rm q}^{\rm G}$ is still positive, though $\sigma_{\rm q}^{\rm PL}$ shifts to a smaller value.
Note that the Gaussian component is almost in the high-mass range for the results obtained by Model \uppercase\expandafter{\romannumeral2}, and we find no evidence that Model \uppercase\expandafter{\romannumeral3} is more preferred than Model \uppercase\expandafter{\romannumeral1}. Therefore, whether the BBHs with primary masses in the Gaussian component have a mass ratio distribution different from the more massive BBHs is inconclusive.

\begin{table*}[htpb]
\begin{ruledtabular}
\caption{Model Comparison Results}\label{BF}
\begin{tabular}{ccccc}
$ln{\mathcal{B}}$   & \multicolumn{4}{c}{$m_1$ Distribution Models and Events Selection}  \\ \cline{2-5}
Models determining the mass ratio distribution    &  PLP \& full  &  PLP \& leave  &  BPL \& full  &  BPL \& leave  
\\ \cline{1-5}
Half Gaussian ($\sigma_{\rm q}$)&0 &0 & 0& 0\\
Model \uppercase\expandafter{\romannumeral1} &2.9&2.5&5.2&5.3\\
Model \uppercase\expandafter{\romannumeral2} ($\sigma_{\rm knot}$=0.5) &4.0&3.5&3.5&5.0\\
Model \uppercase\expandafter{\romannumeral2} ($\sigma_{\rm knot}$=1) &4.2&4.1&4.3&5.5\\
Model \uppercase\expandafter{\romannumeral3} &2.5& 1.5&-&-\\
\end{tabular}
\tablenotetext{}{{\bf Notes.} Here `PLP' and `BPL'  are the abbreviations for PowerLaw Peak and Broken PowerLaw, and `leave' means the case when we leave out GW190412 for analysis. In each case, the values of $ln{\mathcal{B}}$ are relative to the evidence of the model with a half-Gaussian mass ratio distribution.}
\end{ruledtabular}
\end{table*}

\section{Conclusion and Discussion}\label{sec:discussion}

We have investigated the population of BBHs with some parameterized / semiparameterized models to adequately address some potential features in the mass ratio distribution. 
With the current coalescing BBH sample \citep{2021arXiv211103606T}, we first conclude that the mass ratio distribution is not invariable in the whole mass range but varies with the primary mass.
This feature may support the fact that the BBHs observed by LIGO/Virgo/KAGRA may come from not only one evolution process. 
Then we conclude that BBHs in the higher-mass range are more likely to be equal-mass systems than those in the lower-mass range, and the demarcation point $m_{\rm cut}$ may lie between $25M_{\odot}$ and $30M_{\odot}$. Consequently, we predict that asymmetric events are more likely to emerge in the lower-mass range. 
Note that we have excluded GW190814 in our analysis, and our conclusion will be strengthened if we include GW190814, which is potentially located in the lower-mass range.
{ However, from the current observations, we cannot conclude yet whether or not there are additional structures or not in the lower-mass range (see the Appendix \ref{app_lowmass} for the details of the analysis).}
In addition to the common-envelope evolution, some formation and evolution channels that have a stronger preference for producing symmetric BBHs, such as chemically homogeneous evolution \citep{2016MNRAS.460.3545D,2016A&A...588A..50M} and dynamical assembly \citep{2016PhRvD..93h4029R,2017MNRAS.467..524B,2019MNRAS.486.5008A}, have been proposed. BBHs from chemically homogeneous evolution are expected to have total masses above $55M_{\odot}$; meanwhile, for the dynamical assembly in the nuclear star clusters, the heavier BHs are more likely to merge \citep{2022PhR...955....1M}. Therefore, our finding that BBHs in the higher-mass range are more likely to be of equal mass is in concert with the predictions resulting from these formation channels. 

We also find that the BBHs with primary masses in the Gaussian component may have a stronger preference for equal mass than the BBHs in the power-law component. 
This feature in the mass ratio distribution may be associated with the pulsational pair-instability supernovae (PPISNe) \citep{2015ASSL..412..199W,2016A&A...594A..97B}, because the nearly symmetric systems may be the built up of the high-mass BHs created from PPISNe. Additionally, as mentioned above, the chemically homogeneous evolution and the dynamical assembly can also contribute to the nearly symmetric systems in the Gaussian component.
However, currently, we find no evidence that the BBHs in the Gaussian component are more symmetric than those in the higher-mass range, as described above.

We have qualitatively proven that the mass ratio distribution varies with the primary mass in the surveyed mass range, where the asymmetric systems are more likely to emerge in the lower mass range, and the BBHs are more symmetric in the higher-mass range. The parameterized / semiparameterized models considered here are still limited and may not be able to describe the BBH populations completely \citep{2022LRR....25....1M}. 
For example, it would be beneficial to construct a mass-dependent spin model together with the mass ratio distribution, and we leave this for future work. 
Meanwhile, the BBH sample is expected to increase rapidly in the near future \citep{2018LRR....21....3A}. Therefore, with a significantly extended sample, besides better characterizing the mass ratio distribution in the whole mass range, 
new structures may be revealed in the mass spectrum and distribution of spin properties of BBHs, shedding new light onto the BBH formation channels.  

\acknowledgments

{ We thank the anonymous referee for constructive suggestions.} This work was supported in part by NSFC under grants No. 11921003, No. 11703098, and 12073080, the Chinese Academy of Sciences via the Strategic Priority Research Program (Grant No. XDB23040000), Key Research Program of Frontier Sciences (No. QYZDJ-SSW-SYS024). This research has made use of data and software obtained from the Gravitational Wave Open Science Center (\url{https://www.gw-openscience.org}), a service of LIGO Laboratory, the LIGO Scientific Collaboration and the Virgo Collaboration. LIGO is funded by the U.S. National Science Foundation. Virgo is funded by the French Centre National de Recherche Scientifique (CNRS), the Italian Istituto Nazionale della Fisica Nucleare (INFN) and the Dutch Nikhef, with contributions by Polish and Hungarian institutes.\\

\vspace{5mm}
\software{Bilby \citep[version 1.1.4, ascl:1901.011, \url{https://git.ligo.org/lscsoft/bilby/}]{2019ascl.soft01011A},
          PyMultiNest \citep[version 2.11, ascl:1606.005, \url{https://github.com/JohannesBuchner/PyMultiNest}]{2016ascl.soft06005B},
          PyCBC \citep[gwastro/pycbc: PyCBC Release v1.16.14, \url{https://github.com/gwastro/pycbc}]{2019PASP..131b4503B,alex_nitz_2021_5347736}.
          }

\bibliographystyle{aasjournal}
\bibliography{mass_ratio}

\appendix
{ 
\section{Are there additional structures in the lower-mass range?}\label{app_lowmass}
It seems that there are additional structures in the mass ratio distribution in the lower-mass range (i.e., $m_1<30M_{\odot}$) as shown in Fig. \ref{sigma_q}. To investigate this range in more detail, we make some modifications to Model \uppercase\expandafter{\romannumeral1} (hereafter Extended Model \uppercase\expandafter{\romannumeral1}), 
\begin{equation}\label{model4}
\sigma_{\rm q}(m_1 | \sigma_{\rm q}^{\rm low}, \sigma_{\rm q}^{\rm high},m_{\rm cut},\{m_i,f_i\}_{i=1}^{N_{\rm knot}}) = \sigma_{\rm q}(m_1 | \sigma_{\rm q}^{\rm low}, \sigma_{\rm q}^{\rm high},m_{\rm cut}) \text{exp}(f(m_1;\{m_i,f_i\}_{i=1}^{N_{\rm knot}}))
\end{equation}
where $\sigma_{\rm q}(m_1 | \sigma_{\rm q}^{\rm low}, \sigma_{\rm q}^{\rm high},m_{\rm cut})$ is described by Eq. (\ref{model1}). We fix the locations of each knot to be linear in log $m_1$ space of $(5,30)M_{\odot}$, and restrict the perturbation to zero at the minimum knot. Here we consider two settings for the number of knots (${N_{\rm knot}}=5$ and ${N_{\rm knot}}=10$), and three values for $\sigma_{\rm knot}$ (0.2, 0.5, 1.0).

\begin{figure*}
	\centering  
	\subfigbottomskip=2pt 
	\subfigcapskip=-5pt 
	\subfigure[]{
		\includegraphics[width=0.4\linewidth]{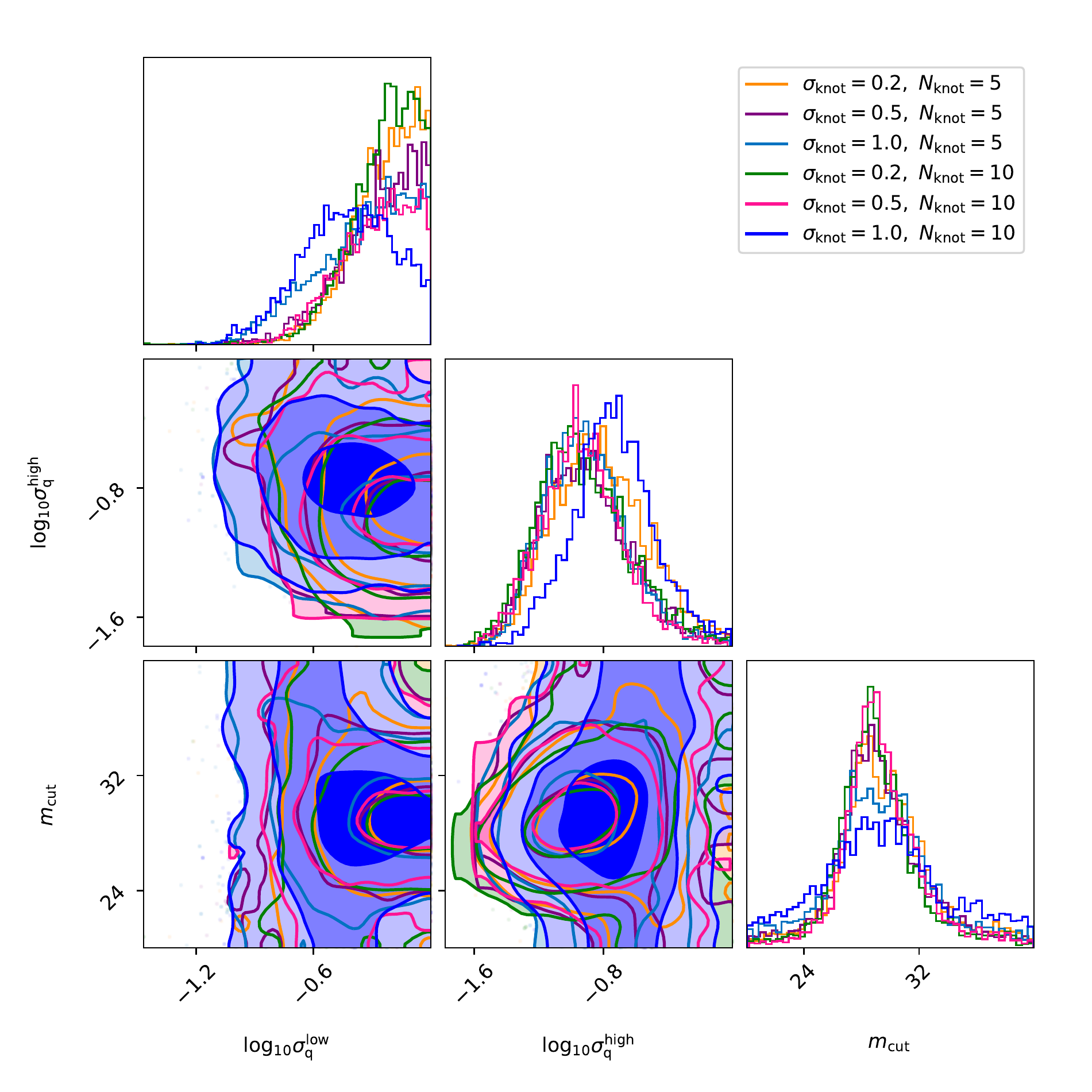}}
	\subfigure[]{
		\includegraphics[width=0.4\linewidth]{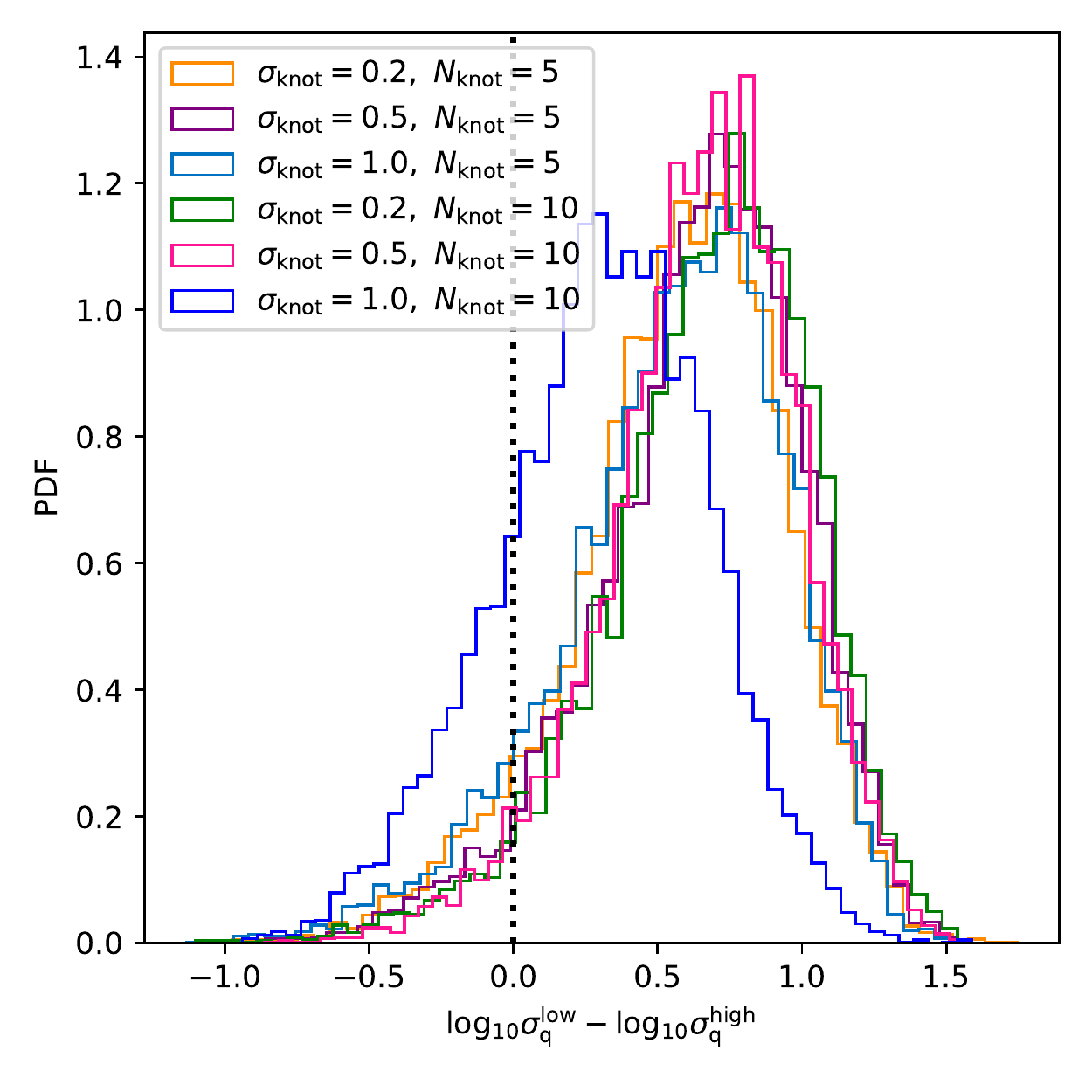}}
\caption{(a): Posterior distributions of $\log_{10}\sigma_{\rm q}^{\rm low}$ ($\log_{10}\sigma_{\rm q}^{\rm high}$) that describe the mass ratio distribution of BBHs in the lower (higher) mass ranges, and the split point $m_{\rm cut}$ obtained by the Extended Model \uppercase\expandafter{\romannumeral1}; (b): Posterior distributions of $\log_{10}\sigma_{\rm q}^{\rm low}-\log_{10}\sigma_{\rm q}^{\rm high}$.}
\label{model4_mcut}
\end{figure*}

Firstly, we find that the values of $\sigma_{\rm q}^{\rm low}$, $\sigma_{\rm q}^{\rm high}$, and $m_{\rm cut}$ obtained in all the cases as shown in Fig. (\ref{model4_mcut}), are consistent with those inferred by Model \uppercase\expandafter{\romannumeral1} as shown in Fig. (\ref{mcut}). Note that a more flexible perturbation function may allow the $\log_{10}\sigma_{\rm q}^{\rm low}$ to support smaller values, like the case of (${N_{\rm knot}}=10$, $\sigma_{\rm knot}=1.0$). 
To find out whether there are additional structures in the mass ratio distribution in the lower-mass range ($m_1<m_{\rm cut}$), we plot the perturbation functions, $f(m_1)$, as shown in the upper row of Fig. \ref{model4_fm}. It shows that the three most apparent perturbations lie at $\sim 9M_{\odot}$, $\sim 13M_{\odot}$, and $\sim 25M_{\odot}$. We find $f(m_1=25M_{\odot})>0$ at 69\%, 84\%, 78\%, and 90\% credibility for the settings of (${N_{\rm knot}}=5$, $\sigma_{\rm knot}=0.5$), (${N_{\rm knot}}=5$, $\sigma_{\rm knot}=1.0$), (${N_{\rm knot}}=10$, $\sigma_{\rm knot}=0.5$), and (${N_{\rm knot}}=10$, $\sigma_{\rm knot}=1.0$), respectively. However, the other two perturbations (i.e., $f(m_1=9M_{\odot})$, and $f(m_1=13M_{\odot})$) are much less significant, as shown in the lower row of Fig. \ref{model4_fm}. 
We find the logarithmic Bayes factors between the Extended Model \uppercase\expandafter{\romannumeral1} and the Model \uppercase\expandafter{\romannumeral1} for all the cases are $\lesssim1$. Therefore, we can not conclude yet that there are additional structures in the mass ratio distribution in the lower-mass range ($m_1<m_{\rm cut}$).
\begin{figure*}
	\centering  
\includegraphics[width=0.95\linewidth]{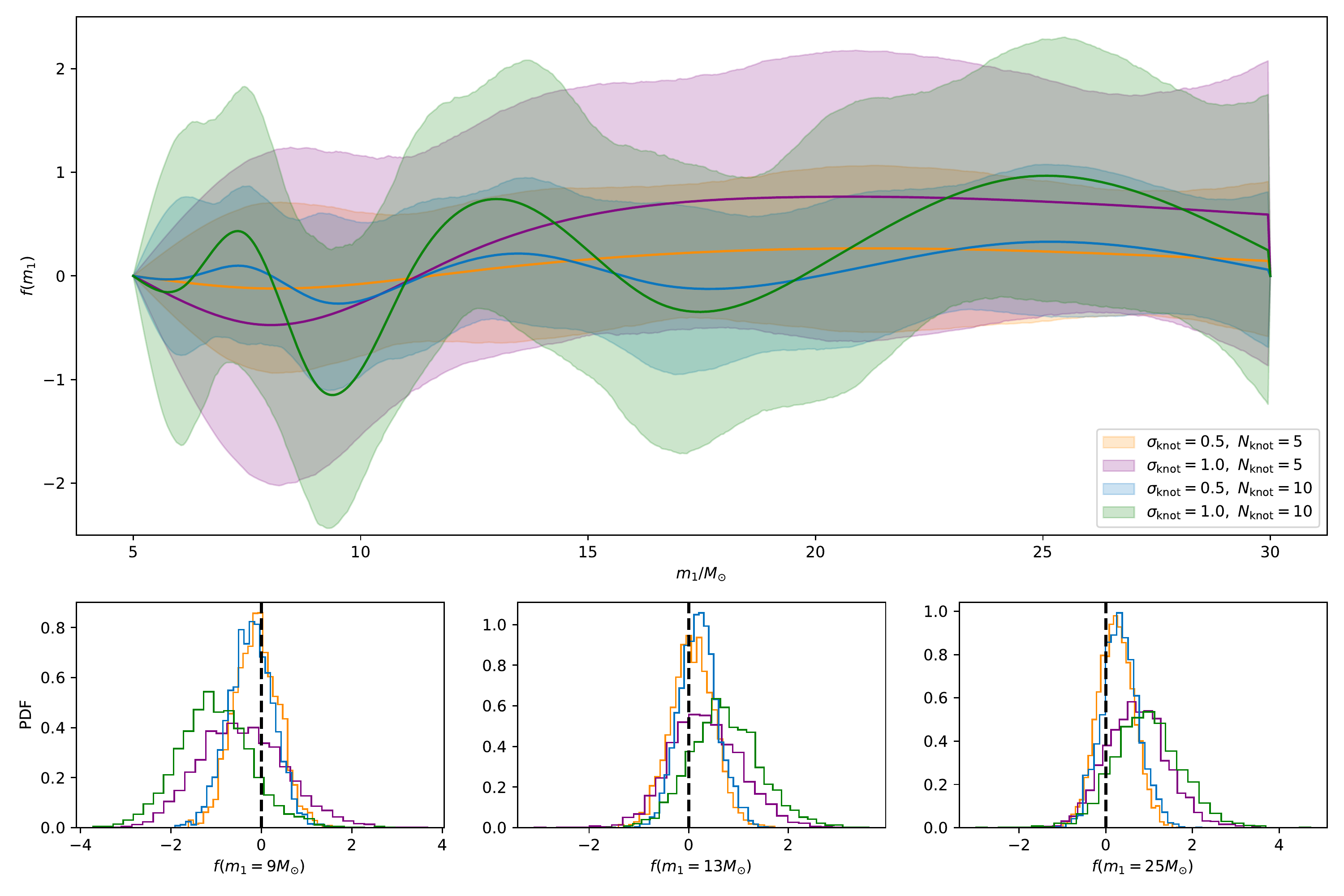}
\caption{The upper row shows the median (solid) and 90\% credible intervals (shaded) of the inferred perturbation functions, $f(m_1)$; the cases for $\sigma_{\rm knot}=0.2$ are not shown here because the $f(m_1)$ is nearly flat .
The lower row shows the posterior distribution of $f(m_1)$ as sliced at the three most apparent inferred perturbations in the posterior which roughly lie at $\sim 9M_{\odot}$ (left column), $\sim 13M_{\odot}$  (middle column), and $\sim 25M_{\odot}$  (right column).}
\label{model4_fm}
\end{figure*}

}

\end{CJK*}
\end{document}